# Boron and Barium Incorporation at the 4H-SiC/SiO$_2$ Interface using a Laser Multi-Charged Ion Source

Md. Haider A. Shaim and Hani E. Elsayed-Ali*

[1]Department of Electrical and Computer Engineering and the Applied Research Center, Old Dominion University, Norfolk, Virginia 23529, USA

*Corresponding author: helsayed@odu.edu; (757)269-5645

**Abstract**

A laser multicharged ion source was used to perform interfacial treatment of the 4H-SiC/SiO$_2$ interface using B and Ba ions. A Q-switched Nd:YAG laser (wavelength $\lambda$ = 1064 nm, pulse width $\tau$ = 7 ns, and fluence $F$ = 135 J/cm$^2$) was used to ablate B and Ba targets to generate multicharged ions. The ions were deflected by an electrostatic field to separate them from the neutrals. The multicharged ions were used for nanometer layer growth and shallow ion implantation in 4H-SiC. Several metal-oxide-semiconductor capacitors (MOSCAP) were fabricated with a combination of B and Ba at the SiC/SiO$_2$ interface. High-low C-V measurements were used to characterize the MOSCAPs. The B interfacial layer reduced the MOSCAP flatband voltage from 4.5 to 0.04 V, while the Ba layer had a negligible effect.

# 1. Introduction

Silicon carbide (SiC) is a material of choice for high power and high temperature operation capable of switching speed at least ten times that of Si and has a very low on-state resistance at high current densities [1]. SiC metal-oxide-semiconductor (MOS) devices are commercially available with blocking voltages up to 1.7 kV and are manufacturable for voltage ratings of over 10 kV [2]. While the carrier mobility in SiC is ~900 $cm^2V^{-1}s^{-1}$, the channel mobility of MOS devices is limited to ~30 $cm^2V^{-1}s^{-1}$ for devices that have been subjected to post-oxidation annealing in NO, which has become the standard process to increase the effective channel mobility [3]. The low channel mobility is due to the presence of high density of $SiC/SiO_2$ interface traps. This causes power losses and reduces device switching speed. In addition to NO anneal, there has been several reports that performed interface treatments using hydrogen [4], phosphorus [5], sodium [6], antimony [7], boron [8], alkali (Rb and Cs), and alkali earth (Ca, Sr, and Ba) interface layers [9]. These treatments were performed individually. While mobility enhancements have been achieved, the threshold voltage stability remains an issue.

The passivation effect and its stability vary significantly depending on the elements used for interfacial treatment. Sodium accumulated at or near the $SiC/SiO_2$ interface increased the channel mobility, but the MOSFET mobility and threshold voltage values were not stable under device operation [10]. Phosphorous increased the channel mobility to ~80 $cm^2/Vs$ but appeared to have a thermodynamic driving force to diffuse throughout the $SiO_2$ forming a phosphosilicate glass, which results in degraded dielectric properties [5]. Boron has been shown to improve the mobility greatly by passivating the interface traps [8]. Boron may also be effective in reducing the interfacial stress, and thus in reducing the trap density, because $B_2O_3$ is known to be a network former that reduces the network connectivity of $SiO_2$ [11]. X-ray photoelectron spectroscopy (XPS) has shown that B reduced the size of carbon clusters at the SiC/SiO$_2$ interface [12]. High-resolution transmission electron microscopy (TEM) and spatially-resolved electron energy-loss spectroscopy (EELS) studies showed that B accumulates in a layer <3 nm at the SiC/borosilicate glass interface and formed a trigonal bonding configuration, softening the oxide and reducing stress at the 4H-SiC interface [13]. This study also showed that P induces roughness, on a scale of hundreds of nm, at the SiC/phosphosilicate interface [13]. Barium has been shown to result in improvement in mobility and threshold voltage stability [9]. The ionic nature of Ba bonding provides more freedom in bond directionality compared to the covalent SiO$_2$ bond which could result in effective carbon removal [9]. Further studies of NO and Ba passivation of the SiC/SiO$_2$ interface showed that Ba remained as a highly localized layer without introducing strain in the interface [14]. For the NO-annealing, N was also concentrated at the interface but, contrary to Ba, introduced significant strain [14].

We used a laser ion source to study the effect of ion deposition accompanied by shallow implantation of B and Ba, individually and together (B + Ba), in 4H-SiC. A laser multicharged ion (MCI) source was used to perform interfacial treatment of the SiC/SiO$_2$ interface using B and Ba ions. Contrary to methods introducing B and Ba by vapor deposition, the laser MCI source causes shallow ion implantation in addition to a nanoscale layer deposition. MOSCAP devices were fabricated with the 4H-SiC surface subjected to low-energy B, Ba, and B+Ba ions causing shallow implantation and a thin layer deposition on the surface. High-low C-V measurements were used to characterize the MOSCAPs. Results show that B at the interface affects the flatband voltage significantly, while the effect of interfacial Ba ions is negligible. B at the SiC/SiO$_2$ interface reduces the flatband voltage from 4.5 to 0.04 V.

## 2. Experimental

### 2.1. The laser ion source

Introduction of B and Ba ions on the surface of the SiC wafers was conducted using an MCI source previously described [15 - 17]. The laser MCI source was modified to allow separation of the ions from the neutrals. The laser ion source used is schematically shown in Fig. 1. It is composed of a 10-cm long plasma expansion chamber and a 66-cm long ion drift tube. The B and Ba targets are grounded during implantation. A grounded mesh anode is placed at a distance of 1 cm from the plasma expansion chamber. When voltage is applied to the plasma expansion chamber, an electric field accelerates the ions after they exist the expansion chamber. If the expansion chamber is grounded, then the ions are accelerated by the double-layer potential at the plasma-vacuum interface with no acceleration aside from that gained in the plasma. Sets of X-Y deflection plates are placed to deflect the ion towards the sample location away from the path of the neutrals. The deflection plates have a length of 2 cm and width of 1 cm. This configuration separates the ions from the neutrals so that ions can be implanted without neutral deposition on the target. A linearly movable sample holder that is used to move the SiC sample into the ion beam. FC-1 (with an area of 20 cm$^2$) is placed at the end of the drift tube, and a small linearly movable FC-2 (with an area of 1 cm$^2$) is placed across the drift tube and is used to detect the deflected ions. The distance from the B and Ba targets to the center of the deflection plates, to FC-1, and to FC-2 are 27, 76, and 42 cm, respectively. A slit (1.5 x 1 cm) is placed near the end of the plasma expansion chamber to limit the ion beam size in the drift tube region and avoid wall sputtering. This arrangement allows for ion implantation and deposition after separating the ions from the neutrals. A Q-switched Nd:YAG ($\lambda = 1064$ nm, $\tau = 7$ ns) is used to ablate the B or Ba targets.

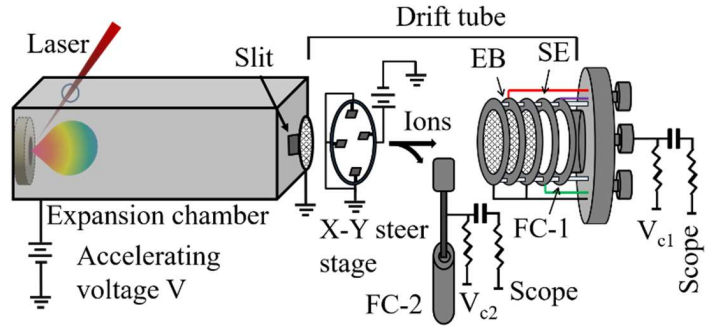

Fig. 1. An illustration of the laser ion source showing the laser irradiating the target, the plasma expansion chamber, drift tube, and the electrostatic time-of-flight energy analyzer. EB electrostatic barrier, SE suppressor electrode, FC-1 Faraday cup-1, and FC-2 is Faraday cup-2.

For B implantation, the laser was focused on the 99.9% pure B target (2-inch diameter, 0.25 inch thick, part number EJTBXXX302A2, from Kurt J. Lesker) to a spot area of $1.3 \times 10^{-3}$ $cm^2$, measured using the scanning knife-edge method at a target-equivalent plane resulting in a laser fluence of 135 mJ/$cm^2$. For Ba implantation, 99.5% pure Ba target (2-inch diameter, 0.25 inch thick, part number EJTBAXX252A4, from Kurt J. Lesker) was used. The laser was operated at 10 Hz repetition rate. The laser-generated plasma plume contains ions with different charge states. The ions gain energy, according to their charge state, due to the voltage developed at the plasma-vacuum interface. Increasing the laser pulse energy increases the generated number of ions, ion energy, and maximum ion charge.

To characterize the ions from the laser plasma with the target and the EC grounded, the voltage bias on the central mesh of the electrostatic barrier (EB) of the three-grid energy analyzer was incrementally increased from 0 V to a voltage that resulted in complete suppression of the ions detected by the FC-1. Fig. 2(a) shows the time-of-flight (ToF) signal for B ions with 0 – 150 V applied to the EB, while Fig. 2(b) shows the ToF signal for Ba ions with 0 – 250 V applied to the EB. The laser fluence used to ablate the B and Ba targets was 135 J/cm$^2$. The applied barrier voltage stops the singly charged ions with kinetic energy lower than the barrier voltage. Whereas, higher charge state ions lose kinetic energy according to their charge state. The retarding field affects each ion charge according to its charge state. The temporal separation of the different ion charges results in a reduction in the amplitude of the ion pulse throughout its temporal width. From Fig. 2(a), we observe that ~50% of the B ions generated with a laser fluence of 135 J/cm$^2$ are retarded by a potential of 25 V. More than 90% of the B ions are retarded by a potential of ~150 V. For Ba and using the same laser fluence, Fig. 2(b) indicates that ~50% of the ions are retarded by a potential of 50 V while 250 V retards ~80% of the ions. This shows that, for a laser fluence of 135 J/cm$^2$, the double-layer potential formed in Ba plasma is higher than that for the B plasma causing higher energy/charge for the Ba ions. The bumps observed in the B ion signal in Fig. 2(a) are due to the contribution of ions of separate charges to the ToF signal. The low atomic number of B ($Z = 5$) allows for ion separation during their ToF to be observed even without applying external ion acceleration. For Ba ($Z = 56$), the expected larger number of ion charges generated result in a smooth ToF signal. Also, comparison of the B and Ba ion signals show that the ion charge from Ba is ~13 times that of B.

For B, we previously reported on the maximum charge of the ions and the ion charge state composition obtained through separating ions with different charge states by applying voltage to the target, which is electrically connected to the expansion chamber [15]. This voltage sets up an accelerating electric field between the end plate of the expansion chamber and the grounded mesh [15, 18].

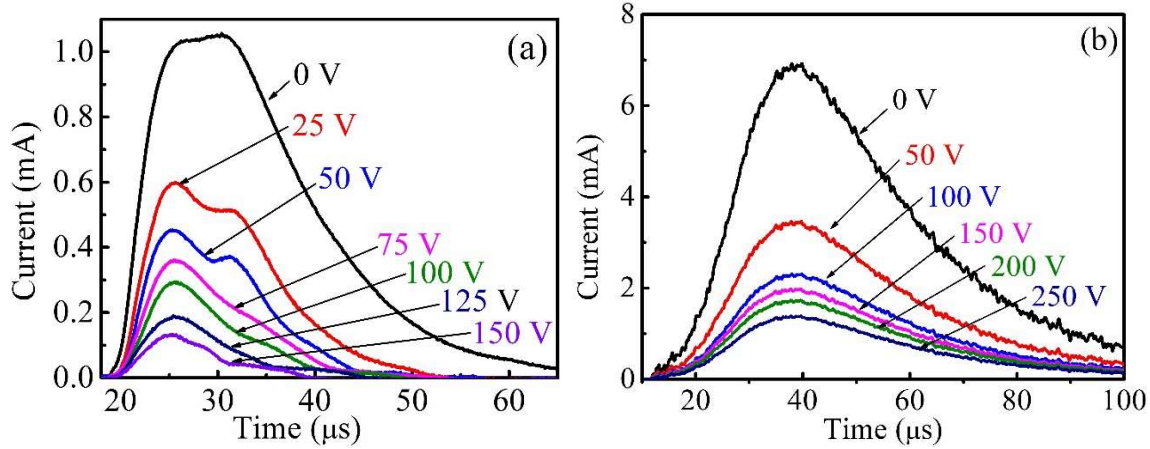

Fig. 2. Ion time-of-flight signal from target irradiated by a laser fluence of 135 J/cm² with different applied barrier (EB) voltages, (a) B target, (b) Ba target. The B and Ba targets were grounded.

The ions reach the detector as a bunch containing different ion states. The Faraday cup FC-2, used to detect the deflected ions, was biased at −70 V, similar to FC-1. The deflecting plates deflect the ions out of the neutral beam path, which is defined by the opening of the expansion chamber and FC-1. Fig. 3 shows the ToF signals for B ions detected by both FC-1 and FC-2 for a laser fluence of 135 J/cm² when the target and the expansion chamber are grounded. For B, the total ions detected by the FC-1 is ~17 nC, when the deflecting plates are grounded. The deflection of ions due to the electric field of the deflection plates can be calculated from $S = \frac{V_d L}{2 d V_{acc}} (D + \frac{L}{2})$, where $S$ is the deflection of an ion from the center of the two plates at a distance $D$ from the end of the deflection plates, $V_d$ is the deflecting voltage, $L$ is the deflection plate length, $d$ is the separation between the plates, and $V_{acc}$ is the ion accelerating voltage. In our experiment, $D = 6.5$ cm, $V_d = 150$ V, $L = 2$ cm, $d = 2$ cm, and $V_{acc} = 150$ V. Ions with different charges are accelerated to a kinetic energy that is proportional to their charge. When a voltage is applied across a set of deflection plates, the ions with similar energy-to-charge ratio are deflected by the same angle. At the location of FC-2 (3.75 cm away from the axis of the transport tube where the neutrals drift), some of the ions are deflected and are separated from the neutrals. This ion deflection is sufficient to assure that only ions reach FC-2. For a deflection plate voltage of 150 V, the total charge detected by the FC-2 is ~9.50 nC, while ~4.50 nC is detected by FC-1. Since the gap between the deflection plates are smaller than the ion beam cross-sectional area, defined by the rectangular opening of the EC, some of the ions continue undeflected in the drift tube to FC-1.

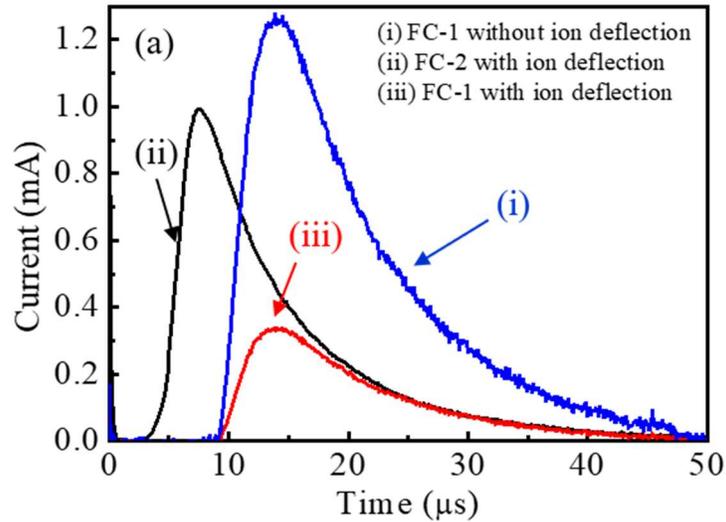

Fig. 3. Boron ion time-of-flight signal generated by a laser fluence of 135 J/cm$^2$. (i) The signals are shown as detected by FC-1 without ion deflection, and (ii) and (iii) for FC-2 and FC-1 with ion deflection using 150 V on deflection plates, respectively.

**3.2 MOSCAP Fabrication**

The substrates were 4H-SiC (orientation <0001>, 4° off axis, dimension 5 x 5 x 0.33 mm, surface roughness < 10 Å, two side polished, Item # SC4HZ0505033S2, from MTI Corporation). Prior to the implantation, the SiC substrate was cleaned using the RCA method to remove the organic, ionic, and metallic impurities followed by HF etching to remove any oxide that may have formed on the top of the SiC substrate, followed by a DI water rinse and N$_2$ dry [19].

To obtain the ion dose, shallow implantation was first conducted using intrinsic silicon (University wafer, ID 2648, Orientation <100>, Resistivity >10,000 Ω-cm) as a target. The MCIs were deflected by the X-Y steering stage towards FC-2 (area 1 cm$^2$). Voltage was applied across the X-deflection plates to bring the center of the ion beam at the substrate position ~2 cm away from the main axis in order to avoid deposition of neutrals on the Si substrate. The ions were detected by FC-2. The B film deposited on the Si substrate was observed by cross-sectional scanning electron microscopy (SEM). The total charge measured is ~7 nC/pulse. The ion bunch detected contained 2.66, 1.61, 1.26, 0.63, and 0.84 nC of $B^{1+}$ to $B^{5+}$, respectively, as obtained by separating the different charges through accelerating the MCIs and detecting the ions ToF signal [16].

Ion deposition and implantation in Si was continued for 10 hours with 10 Hz pulse repetition rate, resulting in a dose of 5.9x10$^{15}$, 1.8x10$^{15}$, 9.45x10$^{14}$, 3.5x10$^{14}$, and 9.45x10$^{13}$ ions/cm$^2$ of $B^{1+}$ to $B^{5+}$, respectively. The total number of ions bombarding the surface during the ion bombardment was ~9x10$^{15}$/cm$^2$. The deposited film thickness was estimated from the total number of ions to be ~11 nm. A similar ion dose calibration was conducted for Ba. Fig. 4 shows the cross-sectional SEM image of B and Ba deposited on the Si substrate. This thickness measurement was used for calibrating the deposition thickness on SiC.

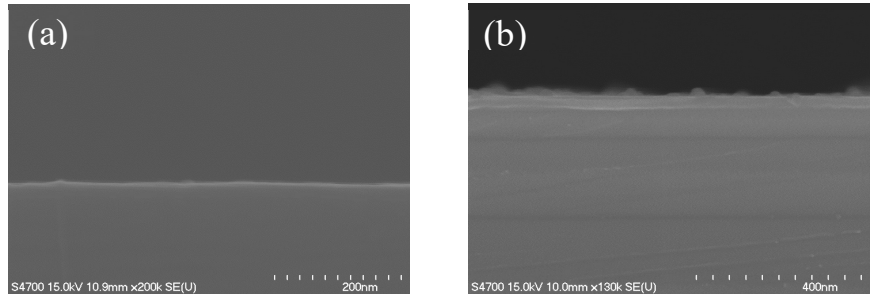

Fig. 4. (a) Cross-sectional SEM image of B film grown on Si showing that for a laser fluence of 135 J/cm$^2$, repetition rate 10 Hz, and irradiation time of 10 hours, ~10 nm film of B was deposited. (b) Cross-sectional SEM image of Ba film grown on Si showing that for a laser fluence of 135 J/cm$^2$, repetition rate 10 Hz, and irradiation time of 1 hour, ~40 nm film of Ba was deposited.

After ion deposition of B and Ba, which also results in shallow ion implantation, a high temperature vacuum tube furnace (MTI Corporation GSL-1100X) was used to anneal the samples at 950 ˚C for 60 min in Ar. Then, ~47 nm of SiO$_2$ was deposited by sputtering using an ATC Orion-5 magnetron sputtering system (AJA International, Inc., USA). An SiO$_2$ target (EJTSIO2452A2, Kurt J. Lesker, 2.00" diameter, 0.25" thickness and 99.995% pure) was used for sputtering. The SiC samples were heated at 250 ˚C during sputter deposition of the SiO$_2$. The base pressure was 3 x 10$^{-7}$ Torr and deposition pressure was 2.2 x 10$^{-3}$ Torr. An additional O$_2$ flow of 6 ccm was used for reactive sputtering. 200 W RF power was used with a deposition rate of 0.5 nm/s. To calibrate the film thickness growth with time, films were grown for a variable time and their cross-sections measured with SEM. The thickness of the oxide was measured post deposition using an ellipsometer (M2000 J.A. Woollam Co.). After depositing the SiO$_2$ layer, the sample was annealed at 950 ˚C for 30 min in Ar.

The gate contacts were 3 mm diameter, ~150 nm of Al sputtered on both sides of the wafer to complete the MOSCAP. The Al thickness was obtained from a crystal thickness monitor placed at the sample location and used to calibrate the deposition rate prior to depositing the Al gates. A schematic of the structure formed is shown in Fig. 5. High-frequency capacitance-voltage (C-V) characterization conducted using Agilent B1500A Semiconductor Device Parameter Analyzer.

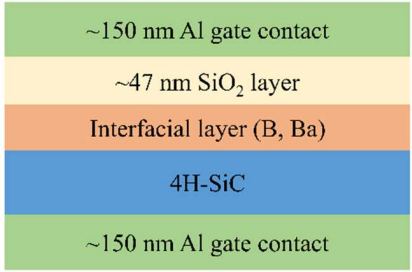

Fig. 5. Diagram showing the structure of the MOSCAP. The interfacial B and Ba layer was formed by multicharged ion deposition/implantation. The SiO$_2$ and Al layers were deposited by magnetron sputtering.

## 3. Results and Discussion

### 3.1. Characterization of the SiC with B surface layer

Fig. 6 shows the Grazing incidence X-ray diffraction (GIXRD) plot for clean 4H-SiC wafer and that after ion deposition/shallow implantation by $1.6 \times 10^{15}$ B ions/cm$^2$, resulting in ~2 nm B film. The only peaks observed are the SiC (111) and (311) peaks. There two peaks are reduced in intensity after B ion deposition. No peaks were observed for B, possibly due to its low coverage and lack or crystallinity.

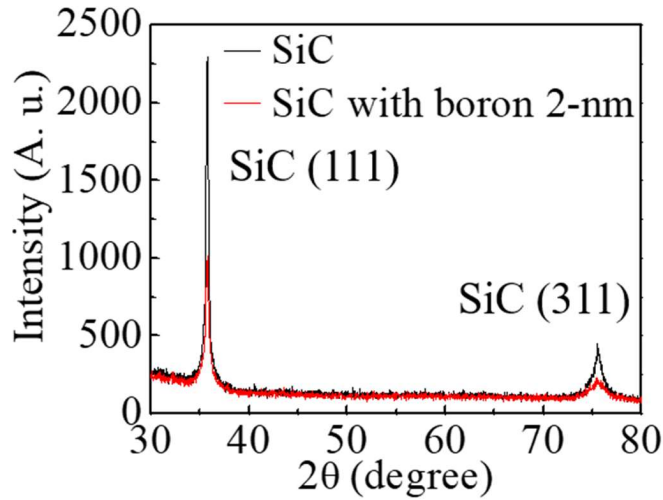

Fig. 6. GIXRD of bare SiC and SiC with ion-deposited 2 nm B layer. The SiC (111) and (311) intensities are reduced after shallow B implantation. The samples were annealed for 60 min. at 950 °C.

The B ion depth profile inside the SiC was checked by time-of-flight secondary ion mass spectrometer (ToF-SIMS) as shown in Fig. 7. The ToF-SIMS was conducted with 25 keV Bi (analysis beam), 0.3 pA current, 50 um x 50 um analysis area, and 1 kV Cs (sputtering beam), 5 nA current, 120 um x 120 um sputter area. When B is implanted in SiC, SIMS detected not only $B^-$, but also adducts including $SiB^-$ and $CB^-$. In Fig. 7, from surface to bulk, depth profiles indicate a very thin oxide layer, a thin B layer, and then SiC. There is also localized distribution of F, Cl and S ion (not shown in the figures). B implanted in the sample is a very shallow, resulting high B concentration on the surface, which alters SiC structure. SIMS quantification is more accurate when doped materials is under 5%. Applying the relative sensitivity factor (RSF) obtained from standard, B concentration in the SiC sample is around $8 \times 10^{21}$ ions/cm$^3$, which is about 16% atomic concentration. This is also why concentration calculated from B, SiB and CB is different.

The Stopping and Range of Ions in Matter (SRIM) simulation was used to check the profile of the implanted B atom [20]. The SRIM simulation was conducted for 10,000 B ions for the ions composed of five energy groups of 150, 300, 450, 600, and 750 eV, with 2000 B ions in each group with energies of 150 to 750 eV, increasing with a step of 150 eV corresponding to $B^{1+}$ to $B^{5+}$. For the simulated 10,000 B ions, the ion range in SiC was 3.1 nm with a straggle of 2.0 nm.

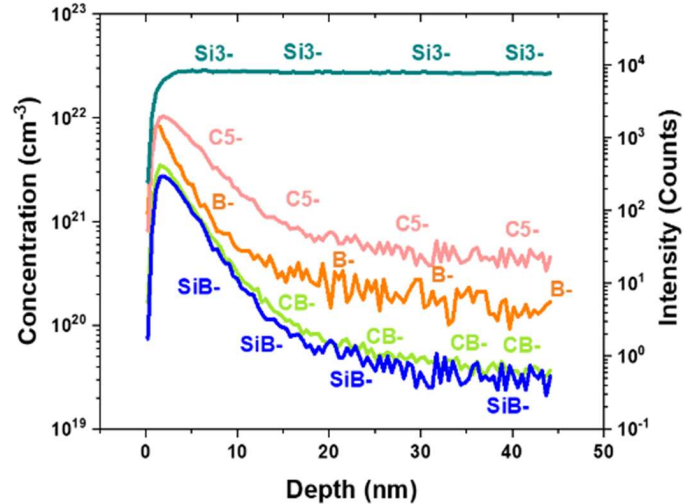

Fig. 7. ToF-SIMS data showing the depth profile of the boron implanted SiC.

**3.2 MOSCAP characterization**

To study the effect of interfacial B and Ba on SiC MOSCAP performance, six MOSCAPs with an identical structure but different $SiC/SiO_2$ interface conditions were prepared. The samples prepared had a $SiC/SiO_2$ interfaces as follows: Bare SiC, 2-nm B, 2-nm Ba, 1-nm B + 1-nm Ba, 2-nm B + 2-nm Ba, and 1-nm Ba + 1-nm B. All samples were coated by ~47 nm $SiO_2$ and the samples were then annealed at 950 ˚C for 30 min and the Al contacts were sputtered on both sides to for the MOSCAP contacts.

High-low C-V characterization was performed by measuring the capacitance from a small amplitude (10 mV) high-frequency and low-frequency AC signal centered around the DC voltage applied to the gate. In the high-low C-V characterization, the voltage is swept from the accumulation to the depletion region of the MOSCAP. In the accumulation region, the electron traps are filled since the conduction band edge is below the Fermi level. The metal Fermi level increases with decreasing DC voltage applied to the MOSCAP gate causing bending of the semiconductor bands upwards [21]. The AC signal is shifted by the DC voltage. The electrons cannot respond to the high frequency by moving into and out of the traps with the voltage change. However, for low frequency C-V characterization, the electrons respond to the voltage and move into and out of the traps as the voltage changes. This electron response creates the differences in charge resulting in higher low frequency capacitance than observed for the high frequency measurement [22].

High-frequency capacitance-voltage (C-V) was characterized at a frequency of 100 kHz at room temperature for MOSCAP with difference B and Ba ion deposition thicknesses as shown in Fig. 8. With increasing B at the interface, the oxide thickness increases, as evident by the reduction of the oxide capacitance, and the C-V curve becoming increasingly left-shifted. While with the increase of Ba dose, the oxide thickness increases with the shift in the C-V curve negligible. Under the flat band condition the effective interfacial charge causes the deviation of the experimental flat-band voltage. The left shift of the C-V curve is due to the interfacial charge carrier change with the doping.

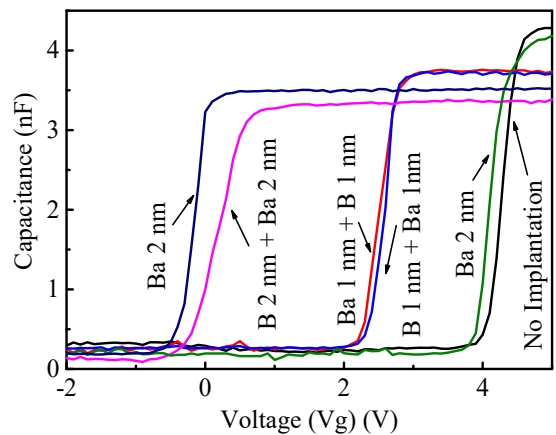

Fig. 8. 100 kHz C-V at room temperature n-type MOSCAP with different B and Ba surface layers.

The flat-band voltage is calculated from the C-V curve using [22]:

$$C_{FB} = \frac{C_{OX} \cdot C_{sFB}}{C_{OX} + C_{sFB}} \quad (1)$$

where $C_{OX}$ is the oxide capacitance and $C_{sFB}$ is the semiconductor surface capacitance in F/cm$^2$, and can be calculated as:

$$C_{sFB} = \frac{\varepsilon_s \varepsilon_o}{L_D} \quad (2)$$

where $\varepsilon_s$ and $\varepsilon_o$ is the dielectric permittivity of the semiconductor and the vacuum in F/cm. $L_D$ is the Debye's length in cm:

$$L_D = \sqrt{\frac{kT\varepsilon_s\varepsilon_s}{q^2 N_D}} \quad (3)$$

where $k$ is the Boltzmann constant (J/K), $T$ is the temperature in K, $q$ is the electron charge in C, $N_D$ is the doping concentration in cm$^{-3}$. The doping concentration can be calculated from the C-V curve. Using the slope $(d/dV).(1/C^2)$ of the linear part of the $(1/C^2)$ versus $V$ characteristics and the $N_D$ value is calculated by [22]:

$$N_D = \frac{2}{q\varepsilon_s\varepsilon_0} / (|slope| \cdot A^2) \quad (4)$$

where $A$ is the gate area in cm$^2$.

The change in the flat-band voltage with varying implantation is given in Table 1. We observe that interfacial B affects the flat-band voltage significantly while the effect of interfacial Ba is negligible.

Table 1. Flat-band voltage with the different interfacial B and Ba.

| SiC/SiO$_2$ interface | Flat-band voltage (V) |
|---|---|
| Bare SiC | 4.5 |
| Ba 2-nm | 4.4 |
| Ba 1-nm + B 1-nm | 2.8 |
| B 1-nm + Ba 1-nm | 2.7 |
| Ba 2-nm + B 2-nm | 0.2 |
| B 2-nm | 0.04 |

High-low C-V measurement of bare SiC and 2-nm B implanted MOSCAP are shown in Fig. 9. For high and low frequency, the 1 MHz and 1 kHz signals are used. From Fig. 9, we observe that the deviation of the high and low C-V curve is increased with the interfacial B, indicating the possibility of interface trap density increase.

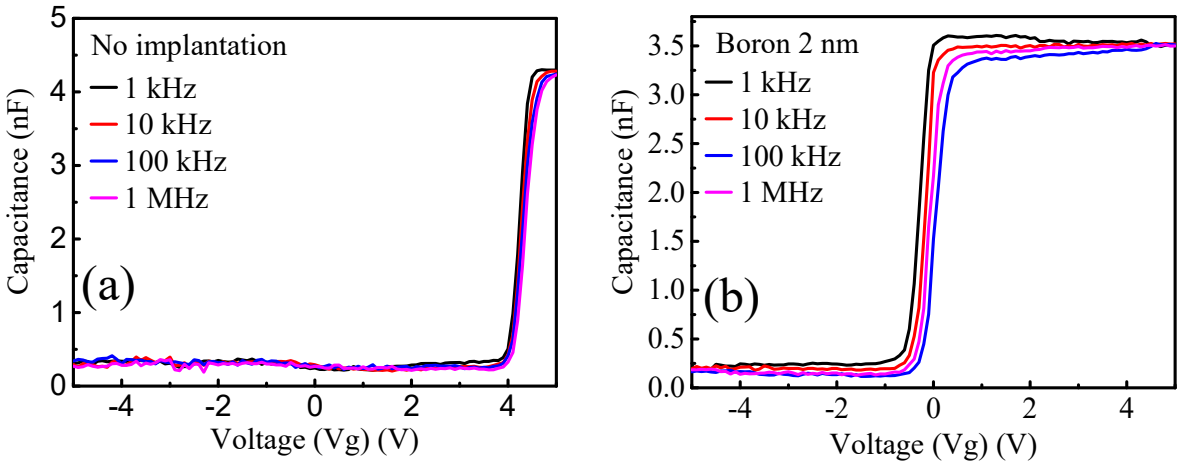

Fig. 9. High-Low C-V curve for bare SiC (a) and 2-nm dose boron implanted SiC (b).

## 4. Summary

Incorporation of B and Ba in SiC/SiO$_2$ interface was performed by ion deposition using a laser multicharged ion source. Different MOSCAPs, with varying B and Ba layers, were fabricated and characterized using high-low CV method. According to SRIM simulation, for ion energy of 150 eV/charge, the ions are implanted up to ~50 Å (FWHM). B shallow implantation/deposition with energy of 150 eV/charge and layer 2-nm thick reduced the flat-band voltage from 4.5 V to 0.04 V, while the effect of Ba implantation was negligible.

**Funding:** This material is based upon work supported by the National Science Foundation, USA, under Grant No. MRI-1228228 and 2000174.

**Conflicts of interest/Competing interests:** The authors declare no conflict of interest.

**Availability of data and material:** All original data and images are available.

**Code availability:** N/A